\documentstyle[12pt,epsf]{article}
\begin{document}
\baselineskip=17pt

\title{\bf Double-beta decay matrix elements \\
\bf for  $^{76}Ge$} 

\author{S. Stoica$^a$, H.V. Klapdor-Kleingrothaus$^b$ \\
a) \it  National Institute of Physics and  Nuclear Engineering, \\
\it P.O. Box MG-6, 76900-Bucharest, Romania\\
b) \it Max-Planck-Institut f$\ddot{u}$r Kernphysik, W-6900 Heidelberg,\\ 
\it Germany} 
\maketitle
\pagestyle{empty}

\vskip.5cm

\abstract{ 
Double-beta decay matrix elements (ME) for  $^{76}Ge$ are calculated 
with different quasi random phase approximation (QRPA)-based methods. 
First, the ME for the two-neutrino  mode are 
computed  using two choices for the single particle (s.p.) basis: 
i) $2-4\hbar\omega$ full shells and ii) $3-4\hbar\omega$ full shells. 
When calculated with the renormalized QRPA (RQRPA) and
full-RQRPA their values are rather dependent on the size of the single 
particle basis used, while calculated  with 
proton-neutron QRPA (pnQRPA) and second-QRPA approaches such a dependence 
was found to be small. The Ikeda sum rule was well fulfilled 
within pnQRPA for both choices of the s.p. basis and 
with a good approximation within second-QRPA, while 
the RQRPA and full-RQRPA methods give deviations up to 21\%. 
Further, the ME for the neutrinoless mode are calculated 
with the pnQRPA, RQRPA and full-RQRPA methods. They all give 
close results for the calculation with the smaller basis (i), 
while for the larger basis (ii), the results differ significantly
either from one method to another or within the same method. 
Finally, using the most recent experimental limit for the 
$0\nu\beta\beta$ decay half-life of $^{76}Ge$ a critical discussion 
on the upper limits for the neutrino mass parameter 
obtained with different theoretical approaches is given.

\vskip0.4cm
\noindent
{\bf Pacs: 21.60.Jz} (Hartree-Fock and random-phase approximations).\\
{\bf Pacs: 23.40.Hc} (Relation with nuclear matrix elements and nuclear 
structure)\\
{\bf Pacs: 23.40.Bw} (Weak interaction and lepton (including neutrino) aspects)
\newpage

\baselineskip=20pt

\section{Introduction}

Since the nuclei which undergo a $\beta\beta$ decay are generally rather far 
from the  closed 
shells, the QRPA-based  methods  have  been extensively employed for computing
ME involved in the theoretical description of this process 
\cite{[HAL67]}-\cite{[BKZ99]}.  Moreover, in spite of the recent progress of the  
shell-model and/or Monte-Carlo shell model techniques
\cite{[CNR96]} these methods also 
remain, at least for the next future, the only available for treating nuclear 
systems which are far away from the closed shells. The pnQRPA \cite{[HAL67]} was 
the first adaptation of the standard  QRPA for nuclear charge-changing 
processes. One of its most important achievements was after the pioneering work of \cite{[GK86]} the success in  
explaining the suppression mechanism of the two-neutrino double beta 
($2\nu\beta\beta$) decay ME \cite{[VOG86]}-\cite{[STAU90]}, reducing thus 
the large discrepancy existing until that moment between the  
theoretical and experimental  $\beta\beta$ decay half-lives.    
However, this method faces the problem of a strong dependence of
these ME on the renormalization of the particle-particle component of the 
residual interaction. Namely, if one represents the ME as function of the 
particle-particle interaction strength (usually denoted by $g_{pp}$), one observes 
that they decrease rapidly and cross through zero in a 
region of physical values of this constant, making the task of fixing it  
adequately difficult. To overcome this problem several further developments of 
this method have been advanced during the recent past. We remind here: the 
appropiate treatment of the particle-number non-conservation \cite{[CFS91]}-\cite{[KRM93]}, 
the inclusion of the 
proton-neutron pairing \cite{[CHE95]}, the double commutator method \cite{[GV92]}, 
\cite{[SUH93]},  computation of the transitions to excited final states 
\cite{[SUH93]}-\cite{[STO94]}, \cite{[SCW97]} as well as the development of approaches 
going beyond the quasi-boson approximation \cite{[RAD91]}, \cite{[STO95]}, \cite{[TOI95]}, 
\cite{[SSF96]}, \cite{[BKZ99]}. 
At this point it is worth mentioning a nice feature of these higher-order QRPA 
approaches:  the like- and unlike-nucleon residual interactions appear both in the next 
higher-order terms beyond pnQRPA, obtaining thus a more realistic picture of the 
competition between  them in producing a $\beta\beta$ decay. 
As a result, calculated with these methods the ME become more stable against 
$g_{pp}$ and the RPA break-down point is shifted towards the region of 
un-physical values of this constant. This is why, the further improvement of such 
approaches  seems to be the most promissing line of development for treating the 
nuclear ME involved in the $\beta\beta$ decay process. 

The first method which has included 
higher-order terms beyond pnQRPA was developed in \cite{[RAD91]} and further, 
applied with some modifications in \cite{[STO94]}-\cite{[RDF]}, \cite{[RS96]}. In this 
approach the extension of the pnQRPA was done using a boson expansion of both the phonon 
operators and transition $\beta^{\pm}$ operators and retaining the next order 
in this expansion beyond the quasi-boson 
approximation (QBA). Also, this method allowed, for the first time, the computation of  
$\beta\beta$ decay rates to excited final states. An alternative approach 
for extending pnQRPA is based on the idea of replacing the uncorrelated QRPA ground state 
(g.s.) by a correlated g.s., in the calculation of the expectation value of the commutator 
of the two bifermion operators involved in the derivation of the QRPA equations. 
The expectation values of the number operator in the QRPA correlated g.s. are 
introduced in the quasi-boson 
commutators of the pair operators and this leads to a renormalization of the QRPA forward-
and backward-going  amplitudes. This method (called RQRPA) was first developed in refs. 
\cite{[HAR64]}-\cite{[ROW68]} for the standard QRPA and adapted later on for 
charge-changing processes in ref. \cite {[TOI95]}. Within the RQRPA a stabilization of 
the ME against $g_{pp}$ and a shift of the RPA break-down point towards larger (un-physical)
 values of this constant are also observed. However, this method has a main  inconvenience 
consisting in an undesirable violation of the Ikeda sum rule (ISR). Some 
refinements in the way of calculating the averages of the quasiparticle number operator 
are proposed \cite{[SSF96]}-\cite{[CHH97]}, but they result in a rather small reduction of 
the violation.     

In this paper we want to make a study of the  
$\beta\beta$ decay nuclear ME of $^{76}Ge$ calculated with different 
QRPA-based methods with the same set of parameters and for both two neutrino 
and neutrinoless modes. The motivation of such a study is given by some  
discrepancies concerning their values which are still found in the 
literature, where similar calculations have been performed.
First, we calculated the nuclear ME involved in the $2\nu\beta\beta$ 
decay mode using the pnQRPA, RQRPA, full-RQRPA and second-QRPA methods. 
One of our goals was to see to what extent the size of the single 
particle basis influences the values of these ME and how one can explain the 
differences between various calculations. A similar study has been made in 
\cite{[SSF96]} but only for the neutrinoless mode. Another point we have focused on was 
to check the Ikeda sum rule (ISR) in the framework of the above mentioned methods. 
Particularly, we would like to compare, under the same conditions of calculation 
(i.e. same parameters and s.p. basis), the various 
degrees of deviations obtained  with these different methods and give possible 
explanations for the differences.  
Further, the ME for the neutrinoless mode are calculated 
with the pnQRPA, RQRPA and full-RQRPA methods. The results  
are found to be close to each other for all three methods in the case 
we used a smaller s.p. basis (9 levels), while for a larger one (12 levels) the results 
differ significantly either within the same method or from one method to another, 
for the two choices of the s.p. basis. 
Then, using the most recent experimental results for the two-neutrino and  
neutrinoless $\beta\beta$ decay half-lives of $^{76}Ge$ \cite{[HM99]}, 
\cite{[HM20]}, we fixed 
first the $g_{pp}$ constant and then extracted new limits for the neutrino mass 
parameter. Finally, we give a critical discussion on the 
values of this parameter obtained with different theoretical 
methods. 
The paper is organized as follows: in section 2 we will give a short 
comparative description of the QRPA-based methods that we used for the 
calculation. The results are presented in section 3 and 
the section 4 is devoted to the conclusions.    

\section{Formalism}

In the QRPA-based methods one assumes the nuclear motion to be harmonic and the excitation 
QRPA operator may have the following general expression:

$$\Gamma^{m+}_{JM^{\pi}} = \sum_{k,l,\mu\leq\mu^{\prime}}\left[X^{m}_{\mu\mu^{\prime}}(k,l,
J^{\pi}) A^{\dagger}_{\mu\mu^{\prime}}(k,l,J,M) + Y^{m}_{\mu\mu^{\prime}}(k,l,J^{\pi}) 
\tilde{A}_{\mu\mu^{\prime}}(k,l,J,M)\right]
\eqno (2.1) 
$$  
\noindent
Here the summation is taken with $k \leq l$ if $\mu=\mu^{\prime}$. 
$X^m$ and $Y^m$ are the forward- and backward-going QRPA amplitudes and $A, A^{\dagger}$ 
the pair quasiparticle operators coupled to angular momentum J and projection M:

$$ A^{\dagger}_{\mu\mu^{\prime}}(k,l,J,M) = {\cal N}(k\mu ,l\mu^{\prime})
   \sum_{m_k,m_l}C^{JM}_{j_k m_k j_l m_l} a^{\dagger}_{\mu km_k} a^{\dagger}_{\mu^{\prime}lm_l} $$

$$\tilde{A}_{\mu\mu^{\prime}}(k,l,J,M) = (-)^{J-M} A_{\mu\mu^{\prime}}(k,l,J,-M)
\eqno(2.2) $$

\noindent
${\cal N}$ is a normalization constant, 
which is different from unity only in case when 
both quasiparticles are in the same shell \cite{[SSF96]}, $\mu, \mu^{\prime} = 1, 2$ and 
$1\equiv$ protons, $2\equiv$ neutrons.
Using the equation of motion method one can derive the pnQRPA equations which, in the matrix 
representation, may be written as:

$$ 
\left(\begin{array}{cc}
\cal A & \cal B \\    
\cal B & \cal A 
\end{array}\right)_{J^{\pi}} \left( \begin {array}{c} X^m \\ Y^m \end{array} \right) = 
\Omega^m_{J^{\pi}} \left(\begin{array}{cc}
\cal U & 0 \\    
 0 & - \cal U  
\end{array}\right)_{J^{\pi}} \left( \begin {array}{c} X^m \\ Y^m \end{array}
\right) \eqno(2.3) $$

\noindent 
where the matrices ${\cal A}$, ${\cal B}$ and  ${\cal U}$ have the following expressions:

$$ 
  {\cal A}_J (\mu k,\nu l; \mu^{\prime} k^{\prime},\nu^{\prime} l^{\prime}) =               
 \langle 0^{+}_{RPA} \vert \left[ A_{\mu\nu}(k,l,J,M), [\hat{H}, A^{\dagger}_{\mu^{\prime}
\nu^{\prime}}(k^{\prime},l^{\prime},J,M)] \right] \vert 0^{+}_{RPA} \rangle 
$$

$$   {\cal B}_J (\mu k,\nu l; \mu^{\prime} k^{\prime}; \nu^{\prime} l^{\prime}) =               
 \langle 0^+_{RPA} \vert \left[ A_{\mu\nu}(k,l,J,M), [\tilde{A}_{\mu^{\prime}\nu^{\prime}}
(k^{\prime},l^{\prime},J,M), \hat{H}] \right] \vert 0^+_{RPA} \rangle 
\eqno(2.4) $$
 $$
{\cal U} =  
\langle 0^+_{RPA} \vert \left[ A_{\mu\nu}(k,l,J,M), [A^{\dagger}_{\mu^{\prime}\nu^{\prime}}
(k^{\prime},l^{\prime},J,M)] \right] \vert 0^+_{RPA} \rangle \eqno(2.5)
$$

\noindent
Here the $\Omega^m_{J^{\pi}}$ are the QRPA excitation energies for the mode $J^{\pi}$.

Within the pnQRPA the QBA is assumed, i.e. the 
quasiparticle operators $A, A^{\dagger}$ are bosons and satisfy exactly
 the boson commutation relations: 
 
$$
\left[ A_{\mu\nu}(k,l,J,M), A^{\dagger}_{\mu^{\prime}\nu^{\prime}}
(k^{\prime},l^{\prime},J,M) \right] =
$$
$$ {\cal N}(k\mu, l\nu) {\cal N}(k^{\prime}\mu^{\prime},l^{\prime}\nu^{\prime}) 
\left(\delta_{\mu\mu^{\prime}}\delta_{\nu\nu^{\prime}}\delta_{kk^{\prime}}
\delta_{ll^{\prime}}- \delta_{\mu\nu^{\prime}}\delta_{\nu\mu^{\prime}}
\delta_{lk^{\prime}}\delta_{kl^{\prime}}(-)^{j_k+j_l-J}\right) \eqno(2.6) $$

\noindent
 In this way the Pauli principle is violated and this 
is a serious drawback of this method. To improve the situation in the 
RQRPA method the $A$, $A^{\dagger}$ operators are renormalized \cite{[TOI95]},
\cite{[SSF96]}:
 
$$
\bar{A}_{\mu\mu^{\prime}}(k,l,J,M) = D^{-1/2}_{\mu k \nu k^{\prime}J^{\pi}}
~A_{\mu\mu^{\prime}}(k,l,J,M) \eqno(2.7)  
$$

\noindent 
where the $D_{\mu k \nu k^{\prime}}$ matrices are defined as follows:
$$
D_{\mu k \nu k^{\prime}J^{\pi}} = 
 {\cal N}(k\mu,l\nu) {\sl N}(k^{\prime}\mu^{\prime},l^{\prime}\nu^{\prime}) 
\left (\delta_{\mu\mu^{\prime}}\delta_{\nu\nu^{\prime}}\delta_{kk^{\prime}}
\delta_{ll^{\prime}}- \delta_{\mu\nu^{\prime}}\delta_{\nu\mu^{\prime}}
\delta_{lk^{\prime}}\delta_{kl^{\prime}}(-)^{j_k+j_l-J} \right) 
$$
$$
\left[1 - j^{-1}_l \langle 0^+_{RPA} \vert [a^{\dagger}_{\nu l}a_{\nu l^{\prime}}]_{00} \vert 
0^+_{RPA}\rangle - j^{-1}_k \langle 0^+_{RPA} \vert [a^{\dagger}_{\mu k}a_{\mu k^{\prime}}]_{00}
\vert 0^+_{RPA} \rangle \right]  \eqno(2.8) $$

\noindent
By inspecting (2.6) and (2.8) one observes that by this renormalization one goes 
beyond the QBA by taking  into account the next terms in the commutator relations 
of the $A, A^{\dagger}$ operators which are just, essentially, the proton and neutron 
number operators. It is worth mentioning that they are taken into account within RQRPA 
only by their averages on the RPA g.s..
The renormalization of the $A, A^{\dagger}$ operators is further carried onto the 
RPA amplitudes, on the $\cal{A}, \cal{B}$ matrices and on the RPA phonon operator also 
obtaining a renormalization of them: 

$$ \bar{X}^m = D^{1/2} X^m~;~\bar{Y}^m = D^{1/2} Y^m~;~
\bar{\cal{A}}^m = D^{-1/2} {\cal A} D^{-1/2}~;~\bar{\cal{B}}^m = D^{-1/2} {\cal B} D^{-1/2} 
\eqno(2.9) $$

$$\Gamma^{m+}_{JM^{\pi}} = \sum_{k,l,\mu\leq\mu^{\prime}}\left[\bar{X}^{m}_{\mu\mu^{\prime}}
(k,l,J^{\pi}) \bar{A}^{\dagger}_{\mu\mu^{\prime}}(k,l,J,M) + \bar{Y}^{m}_{\mu\mu^{\prime}}
(k,l,J^{\pi}) \tilde{\bar{A}}_{\mu\mu^{\prime}}(k,l,J,M)\right]
\eqno (2.10) 
$$  
\noindent
To calculate $\bar{\cal A}$ and $\bar{\cal B}$ we need to determine the renormalization matrices $D$.
This is done by solving a system of non-linear equations for them by an iterative 
numerical procedure. As input values one can use their expressions in which the averages of the
number operators are replaced by the back-forwarded amplitudes obtained as initial solutions of 
the QRPA equation. 

In QRPA-type methods, before starting the RPA procedure, we need the occupation 
amplitudes (u, v) and the quasiparticle energies, in order to get the 
image of the RPA operators in the quasiparticle representation. This is done by solving the HFB 
equations, which may include, in the general case, both like- and unlike-nucleon pairing. 
When one includes only like-nucleon pairing in these equations, the QRPA procedure described 
above was called RQRPA \cite{[TOI95]}, \cite{[SSF96]}, \cite{[SIM97]}, \cite{[BKZ99]}, while 
when both types of the pairing interaction are included 
it was called full-RQRPA \cite{[SSF96]}, \cite{[SIM97]}. On the other hand, if one takes the $D =1$ 
we get back the QBA and these methods become pnQRPA and full-QRPA, respectively.    

In the second-QRPA method the principle of including higher-order corrections to the pnQRPA 
and restoring partially the Pauli principle is different. Here, the two quasiparticle and 
the quasiparticle-density dipole operators are expanded in a Beliaev-Zelevinski series 
\cite{[BZ62]}:

$$ 
A^{\dag}_{1\mu}(pn) = \sum_{k}\left( {\it A}^{(1,0)}_{k_1}\Gamma^+_{1\mu}(k) + 
{\it A}^{(0,1)}_{k_1}\tilde{\Gamma}^+_{1\mu}(k)\right)
\eqno(2.11) $$

$$
B^{\dagger}_{1\mu}(pn) = \sum_{k_1k_2} \left({\it B}^{(2,0)}_{k_1k_2}(pn)
[\Gamma^{\dag}_1(k_1)
\Gamma^{\dag}_2(k_2)]_{1\mu} + {\it B}^{(0,2)}_{k_1k_2}(pn)[\Gamma_1(k_1)
\Gamma_2(k_2)]_{1\mu}\right) \eqno(2.12)
$$

\noindent
where 
$$ 
B^{\dagger}_{1\mu}(pn) = 
   \sum_{m_k,m_l}C^{JM}_{j_pm_pj_nm_n} a^{\dagger}_{j_p m_p} a_{j_n m_n} 
$$

$$\tilde{B}_{1\mu}(pn) = (-)^{J-M} B_{1\mu}(pn)
\eqno(2.13) $$

The boson expansion coefficients ${\it A}^{(1,0)}$, ${\it A}^{(1,0)}$, ${\it B}^{(2,0)}$, 
 ${\it B}^{(0,2)}$ are determined so that the equations (2.11)-(2.12) 
are also valid for the corresponding ME in the boson basis. 

Further, the transition $\beta^{\pm}$ operators in the quasiparticle representation 
can be expressed in terms of the dipole operators $A_{1\mu}$ and $B_{1\mu}$:
 
$$
\beta^{-}_{\mu}(k) = \theta_k A^{\dag}_{1\mu}(k) + \bar{\theta}_k \tilde{A}_{1\mu}
                    +  \eta_k B^{\dag}_{1\mu}(k) + \bar{\eta}_k \tilde{B}_{1\mu}
$$
$$
\beta^{+}_{\mu}(k) = -\left(\bar{\theta}_k A^{\dag}_{1\mu}(k) + \theta_k \tilde{A}_{1\mu}
                    +  \bar{\eta}_k B^{\dag}_{1\mu}(k) + \eta_k \tilde{B}_{1\mu}\right)
\eqno(2.14) $$

\noindent 
where 

$$ \theta_k = \frac{\hat{j}_p}{\sqrt{3}} \langle j_p \vert \vert \sigma \vert \vert j_n \rangle
U_pV_n; ~\bar{\theta}_k = \frac{\hat{j}_p}{\sqrt{3}} \langle j_p \vert \vert \sigma \vert 
\vert j_n \rangle U_nV_p;~ \hat{j}=\sqrt{2j + 1}$$

$$
 \eta_k = \frac{\hat{j}_p}{\sqrt{3}} \langle j_p \vert \vert \sigma \vert \vert j_n \rangle
U_pU_n; ~\bar{\eta}_k = \frac{\hat{j}_p}{\sqrt{3}} \langle j_p \vert \vert \sigma \vert 
\vert j_n \rangle V_pV_N \eqno(2.15)
$$
\noindent
Using the boson expansions (2.11)-(2.12),  
one also gets expressions of the transition operators beyond the quasiboson 
approximation. Thus, in the second-QRPA method, the higher-order corrections to the 
pnQRPA are introduced not only in the RPA wave functions (by improving the phonon operator 
with additional correlations), but also in the expressions of the $\beta^{\pm}$ operators, 
and the procedure is now more consistent. 
The additional terms will have, of course, an influence on the ME calculation 
of these operators .          

Further, we give the factorized forms of the two-neutrino and neutrinoless $\beta\beta$ decay 
half-lives that we used in our calculations :

$$ \left[T^{2\nu}_{1/2}\right]^{-1} = F^{2\nu} \vert M^{2\nu}_{GT}\vert^2 \eqno(2.16)$$

\noindent 
where $F^{2\nu}$ is the lepton space phase and 
$$
M^{2\nu}_{GT} = \sum_{l,k}\frac{\langle 0^+_f\vert\vert \sigma\tau^-\vert\vert 1+_k\rangle
\langle 1^+_k \vert 1^+_l\rangle \langle 1^+_l \vert \vert \sigma\tau^- \vert \vert 0^+_i\rangle}
{E_l + Q_{\beta\beta}/2 + m_e - E_0} \eqno(2.17) $$

\noindent
In (2.17) l, k denote the two different sets of $1^+$ states in the odd-odd nucleus
obtained with two separate RPA procedures applied onto the g.s. of the initial and final 
nuclei 
participating in the $\beta\beta$ decay. $E_l$ is energy of the $l-th$ intermediate $1^+$ 
state, and $E_0$ is the initial g.s. energy. 

$$ 
\left[T^{0\nu}_{1/2}\right]^{-1} = C_{mm}\left(\frac {\langle m_{\nu}\rangle}{m_e}\right)^2
\eqno(2.18) $$

\noindent
where $\langle m_{\nu}\rangle$ is the effective neutrino mass and  

$$ C_{mm} = F_1^{0\nu}\left(M^{0\nu}_{GT}- \left(\frac{g_v}{g_A}\right)^2 M^{0\nu}_{F} 
\right)^2 = F_1^{0\nu}\cdot (M^{0\nu})^2 \eqno(2.19) $$

\noindent 
$ F_1^{0\nu}$ is the phase-space integral and $M^{0\nu}_{GT}$ and $M^{0\nu}_{F}$
are Gamow-Teller and Fermi matrix elements.      

\section{Results}

\subsection{Two-neutrino double beta decay}

First, we have performed calculations of the nuclear ME involved in the 
$2\nu\beta\beta$ 
decay mode of $^{76}Ge$ using the pnQRPA, RQRPA, full-RQRPA and second-QRPA 
methods. 
For the s.p. basis we used two  choices. We included: i) the 12 levels  
belonging to the full sd, pf and sdg shells, taking thus $^{16}O$  
as core and ii) the 9 levels belonging to the full pf and sdg shells and  
taking thus $^{40}Ca$ as core. 
The single particle energies have been obtained by solving the 
Schr$\ddot{o}$dinger 
equation with a Coulomb-corrected Woods-Saxon potential. For the residual two-body 
interaction there 
was taken the Brueckner G-matrix calculated from a Bonn-OBEP. The quasiparticle 
energies 
and the BCS occupation amplitudes were derived by solving the HFB equation without 
and with 
proton-neutron pairing, separately for the initial and final nuclei, with both
choices of the s. p. basis.  For a complete calculation 
we included in the model space the states with all the multipolarities $J^{\pi}$. 
The renormalization constants were chosen as 
follows: $g_{pp}=1.0$ for all the multipolarities, except the $1^+$ channel for 
which it 
was left as a free parameter, and $g_{ph}=1.0$ for all the multipolarities except 
the 
$2^+$ channel where it was fixed to $0.8$, since for larger values the p-h 
interaction in 
this channel is too strong producing the collapse of the RPA procedure.
The value of all the constants, including those which renormalize the 
pairing interactions are presented in the Table 1.      

\begin{table}
\centerline{   
\begin{tabular}{|cccccccc|}
\hline
 & No. levels& $g_{ph}$ & $g_{ph}^2$ & $g_{pp}^*$ & $g_{pair}^n$ & $g_{pair}^p$ & $g_{pair}^{pn}$ \\
\hline 
$^{76}Ge$ & $9$ & $1.0$ & $0.8$ & $1.0 $ & $1.230 $ & $1.057$ & $2.071 $ \\
 &$12$ & $1.0$ & $0.8$ & $1.0$ & $1.011$ & $1.034 $ & $1.751 $ \\
\hline
$^{76}Se$ & $9$ & $1.0$ & $0.8$ & $1.0 $ & $1.138 $ & $1.214$ & $1.653 $ \\
 &$12$ & $1.0$ & $0.8$ & $1.0$ & $1.031$ & $1.194 $ & $1.502 $ \\
\hline
\end{tabular}}
\caption{Values of the renormalizing constants.}
\label{t1}
\end{table} 

In Fig. 1 we displayed the $M^{2\nu}_{GT}$ (in $MeV^{-1}$) as function of 
$g_{pp}$ calculated with pnQRPA and second-QRPA methods. The two curves for each method 
represent the calculations performed with the two different s.p. basis.
In the figure is also drawn the line representing the ME value corresponding to the 
latest experimental  $2\nu\beta\beta$ decay half-life of the  $^{76}Ge$,
obtained by the Heidelberg-Moscow experiment: $T^{2\nu}_{1/2}=1.55 \times 
10^{21}$ yr (\cite{[HM20]}). 

\vspace*{2cm}
\begin{figure}[ht]
\centerline{\epsfxsize=12cm\epsffile{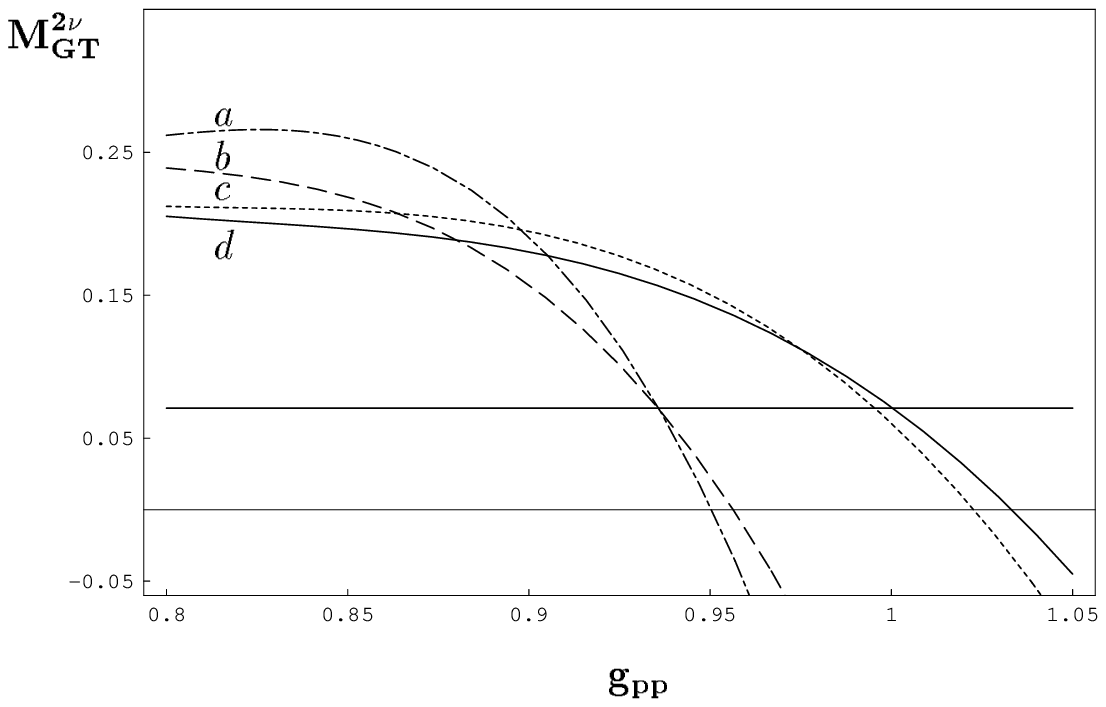}}
\caption{$ a = pnQRPA (12) $~;~$ b = pnQRPA (9) $~;~$ c = second-QRPA (9)$~;
$ d = second-QRPA (12) $}
\label{Figure 1}
\end{figure}

\noindent
As it was already observed in previous calculations \cite{[RAD91]}, \cite{[STO94]} 
the point where QRPA breaks down is pushed to higher values of $g_{pp}$ in the 
framework of the second-QRPA method as compared with the pnQRPA.
The calculation also shows that, within these two methods 
the values of the ME do not depend significantly on the size of the s.p. space, 
especially in the region around the experimental value. The values of $g_{pp}$ 
which fit the best 
this experimental value are: $0.94$ for both calculations performed with pnQRPA and 
$ 0.99$ and $1.01$ for the calculation with 12 and 9 levels, respectively performed with 
second-QRPA.

\vspace*{2cm}
\begin{figure}[ht]
\centerline{\epsfxsize=12cm\epsffile{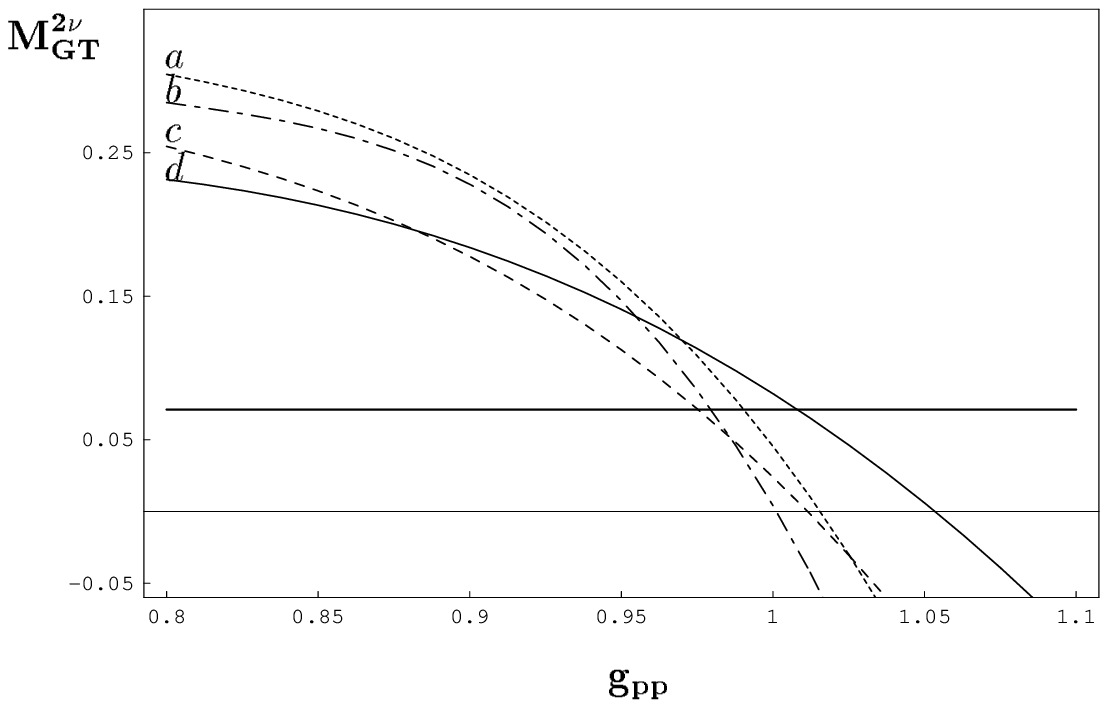}}
\caption{$ a = full-RQRPA(12)$~;~$ b = full-RQRPA(9)$~;~$ c = RQRPA(9)$~~;
~~~$ d = RQRPA(12) $}
\label{Figure 2}
\end{figure}

In Fig. 2 are displayed the same ME but calculated with RQRPA and full-RQRPA 
methods, 
Contrary to the previous calculations, in this case the difference between 
the results
 obtained with different choices of the s.p. space, within the same method, is 
rather large. Indeed at values of $g_{pp}$ 
where $M^{2\nu}_{GT}$ crosses the line representing the experimental result, the values for the 
ME, obtained with the same method, differ from each other by up to $40-50\%$ when the 
two different choices of the s.p. basis are used. The values of 
$g_{pp}$ for the best fits with the experimental value for the ME are: $0.977$; $0.982$ 
in the case of the RQRPA and $0.975$; $1.012$ in the case of the full-RQRPA
for calculations including 9 and 12 levels in the s.p. basis, respectively. The 
differences persist when the s.p. basis was enlarged to 21 states. 
This different behavior of the calculation, obtained with RQRPA and 
full-RQRPA on one side, and with pnQRPA and second-QRPA one the other side, in connection
to the choices of the s.p. basis, reflects the sensitivity of the former methods in 
computing the $M^{2\nu}_{GT}$ ME. One possible source of this sensitivity might have its 
origin in the numerical computation.
Indeed, the self-consistent iteration procedures for solving the full RQRPA equations, 
for all the multipolarities, are very time-consuming and rather slow converging and 
might affect the precision of the calculation. For more reliable calculations 
improved numerical techniques are in our opinion further required.

On the other hand, there are some theoretical arguments which could explain the different 
results for the ME obtained with RQRPA-like methods as compared to the other two. We 
will discuss them later, after having discussed the ISR. 

The ISR
 
$$ S_- - S_+ = \Sigma_m \vert\langle  0^+_{gs}\vert\vert
\beta^-_m\vert\vert 1^+_m\rangle \vert^2- \Sigma_m \vert\langle
0^+_{gs}
\vert\vert \beta^+_m\vert\vert 1^+_m\rangle \vert^2 \eqno(3.1) $$

\noindent
was checked out in the framework of the four methods. 
The results are presented in Table 2, where the percentages of deviation from the correct 
value for each method and choice of the basis are given. The first values in the row 
represent the calculations with a s.p. basis with 12 levels, while the second numbers 
refer to the  same calculation, but with 9 levels.
     
\begin{table}
\begin{center}
\begin{tabular}{|ccccc|}
\hline
 & pnQRPA & RQRPA & full-RQRPA & second-QRPA \\
\hline 
$^{76}Ge$ & $0.23$~~$0.26$ & $20.06$~~$21.34$ & $17.68$~~$17.29$ &$2.7$~~$
3.7$ \\
\hline
$^{76}Se$ & $0.41$~~$0.48$ &$19.66$~~$19.94$ & $17.12$~~$16.91$ 
&$3.13$~~$4.18$ \\
\hline
\end{tabular}
\caption{The numbers represent the deviations (in percents)
from the ISR calculated within the specified methods.
The first (second) numbers in the rows represent the calculation 
with a s.p. basis containing 12 (9) levels, respectively.}
\label{t2}
\end{center}
\end{table} 

One can see, as expected, that within the pnQRPA the ISR is very well fulfilled, while within 
RQRPA and full-RQRPA the deviations are between $17-21\%$. 
One can also observe that within second-QRPA 
the deviations from the ISR are rather small, confirming the result reported earlier in 
refs. \cite{[STO94]}, \cite{[STO95]}, but at that time calculated including only the $1^+$ 
channel and 9 levels in the s.p. basis. We should mention that in the present second-QRPA 
calculation we did not take into account the three boson states which may introduce 
undesirable spurious states in the QRPA space. Including  such states one also 
gets deviations up to $17\%$ from the ISR. 
Looking for some theoretical arguments for a possible explanation 
of the different extent to which the ISR is fulfilled within the RQRPA and 
second-QRPA methods, one finds that  
one reason could be the existence of some inconsistencies of the 
RQRPA related to the way of partial restoration of the Pauli principle.

\vspace*{2cm}
\begin{figure}[ht]
\centerline{\epsfxsize=12cm\epsffile{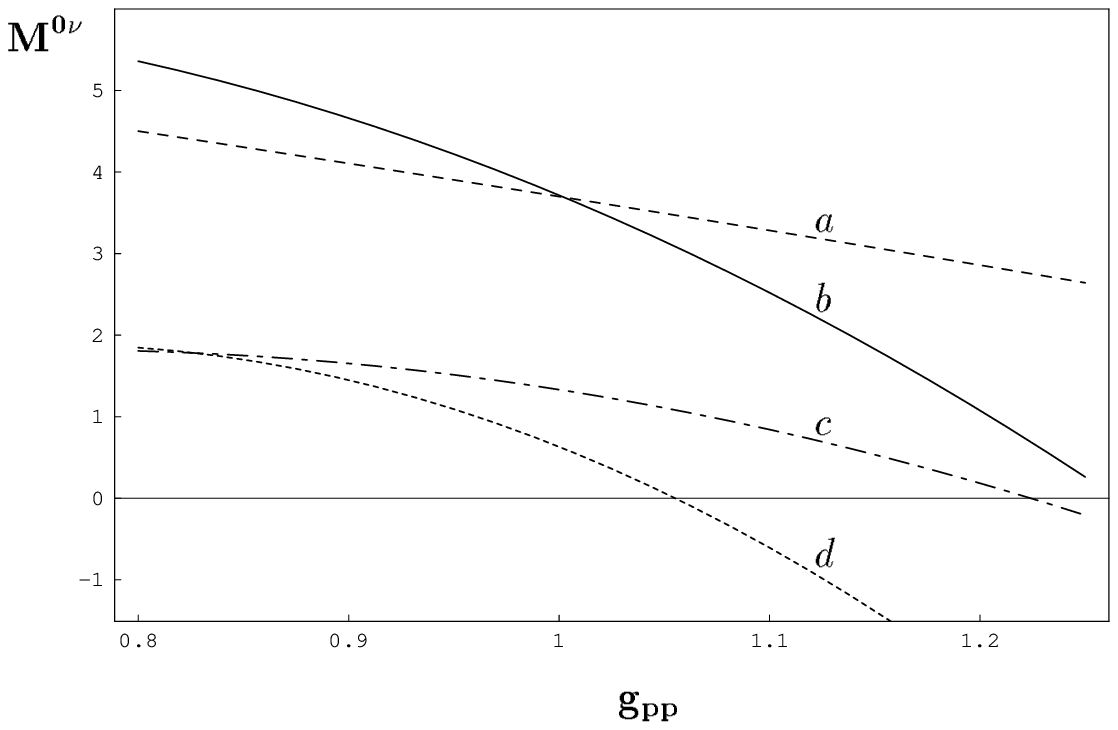}}
\caption{$ a = RQRPA(9) $~;~$ b = pnQRPA(9) $~;~$ c = RQRPA(12) $~;~
$ d = pnQRPA(12)$}
\label{Figure 3}
\end{figure}

Indeed, as we already mentioned in section 2, within the RQRPA method the Pauli principle is 
partially restored for the operators $A$, $A^{\dagger}$, by taking into account the 
averages of the quasiparticle-number operators in their commutator relations. However, 
there is no justification to neglect them in the $B, B^{\dagger}$ operator commutation 
relations. Within the second-QRPA higher order corrections beyond pnQRPA are 
taken into account in the expression of these operators and moreover, such 
corrections are also introduced in the expressions of the $\beta^{\pm}$ 
operators. 
The effect of the additional terms combined with a larger boson space 
(in the second-QRPA the boson space enlarges from one to two boson states) 
reflects in  a 
positive contribution to the ISR. On the other hand it is known that RQRPA 
underestimates the ISR, so this could be one possible explanation why the ISR is better fulfilled within 
second-QRPA. Another possible shortcoming of the RQRPA method is a lack of 
consistency between 
BCS and QRPA levels. While in the BCS still one assumes the g.s. to be the quasiparticle 
vacuum at the level of RQRPA we are dealing with the non-vanishing quasiparticle content 
of the g.s. due to the additional scattering terms taken into account in the commutation 
relations \cite{[BKZ99]}.

\subsection{Neutrinoless double-beta decay}

Further, we have performed a calculation of the neutrinoless ME using pnQRPA, RQRPA and 
full-RQRPA 
methods, also for the two s.p. basis. The $M^{0\nu}$ as function of 
$g_{pp}$ calculated with the pnQRPA and RQRPA are displayed in Fig. 3, while the same ME 
but 
calculated with the full-RQRPA are displayed in Fig. 4. One observes that all the three 
methods 
used for calculation give different values of the ME for different choices of the s.p. 
basis. 
The differences between ME values calculated with 9 and 12 levels included in the s.p. 
basis, within the pnQRPA and RQRPA  methods, are given by factors of 
3 and 2.5, respectively,  while for the full-RQRPA method the difference 
between the two calculations reduces to a factor of about 1.6.  
One also observes, that the values of the ME obtained with the three methods are close to 
each other (3.9-4.1) in the calculation with the smaller basis. When enlarging 
the basis to 21 levels, the result is close to that obtained with the basis with 
12 levels. 
This again reveals the sensitivity of the RQRPA-type methods to the choice of 
the s.p. basis and seems to indicate a possible stabilization of the results 
for larger basis. However, a general conclusion about which basis is better 
to choose is difficult to give until we have not the whole image of a 
similar study performed on several other double-beta 
emitters. Another still open question is what are the results when a similar 
calculation is performed with the second-QRPA.  

\vspace*{3cm}
\begin{figure}[ht]
\centerline{\epsfxsize=12cm\epsffile{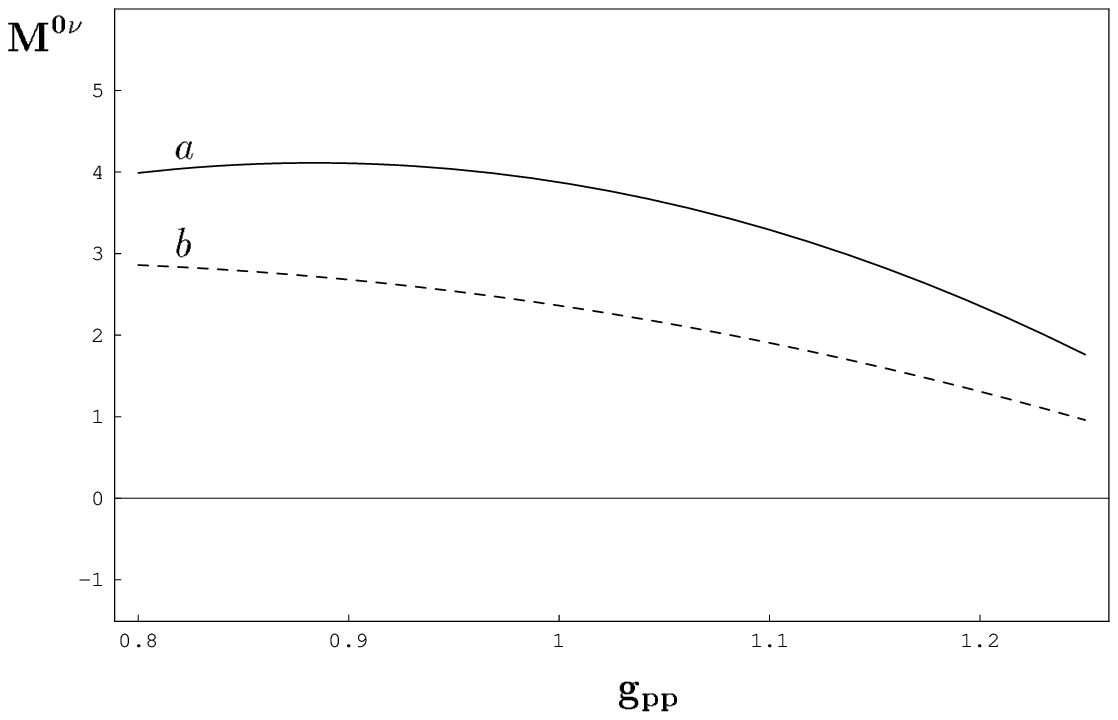}}
\caption{$ a = full-RQRPA(9) $~;~$ b = full-RQRPA(12)$}
\label{Figure 4}
\end{figure}

Finally, using the value of the $g_{pp}$ constant, fixed from the $M^{2\nu}_{GT}$ calculation 
to fit the most recent half-life, i.e. 1.0 (very close to the average value between the 
two second-QRPA and full-RQRPA calculations), and using the most recent experimental limit of 
the neutrinoless mode half-life for the $^{76}Ge$ case (i.e.  
$ > 1.9 \times 10^{25}$ yr (90\% C.L.) reported by the Heidelberg-Moscow 
experiment \cite{[HM20]}) we extract new upper limits for the 
neutrino mass parameter within the full-RQRPA method. We obtain
the values = $0.407$ eV and $0.625$ eV, if we use in the calculation a 
s.p. basis with 9 and 12 levels, respectively. 
In addition, in table 3, besides our
values for the ME and neutrino mass parameter, we present results of other 
calculations found in the literature. For a direct comparison 
between various results we use non-dimenssional values for all the ME
taken from the references indicated in the table. Further, 
using the same phase space factor $F_1^{0\nu} = 6.31 
\times 10^{-15}~yr^{-1} $ \cite{[SC98]} and the same 
half-life reported in Ref. \cite{[HM20]}, 
we extracted upper limits for $<m_{\nu}>$ corresponding to 
all values of the ME, in order to have a more complete image of the evolution 
of the theoretical predictions. Performed with pnQRPA, the older 
results are very similar, although they were calculated by 
different groups and with different numerical codes and parameters 
(\cite{[SMK90]}, \cite{[TOM91]}). One also observes that their values 
are larger by a factor of about two than the values obtained by 
using the recent extensions of pnQRPA, RQRPA and full-RQRPA.
However, it should be kept in mind that these last approaches do not 
fulfill the ISR and thus lead apriori to too small M.E i.e. too large 
neutrino mass limits .
Our calculations performed with the larger s.p. basis 
give M.E. rather close to those of Ref. \cite{[SPV20]} 
were corrections due to the nucleon currents such as weak magnetism and
pseudoscalar couplings to the amplitude of $0\nu\beta\beta$ have been
taken into account. The use of the smaller s.p. basis yields a value for 
the M.E. which close to the earlier approaches \cite{[CFT87]}, 
\cite{[SMK90]}, \cite{[TOM91]}. It should be pointed out that corrections due 
to nucleonic currents mentioned above were not make in the present study.

On the other side there are the values of the M.E. calculated
with the shell-model in Refs. \cite{[HAX84]}, \cite{[CNR96]} which differ 
from each other by a factor of 3. However, in our opinion, these 
calculations performed with the shell model are not, at present, reliable enough. This has been stressed also by \cite{[FAE99]}. 
In ref. \cite{[HAX84]} the calculations were performed with a rather crude
shell-model code in a weak coupling approximation, at the computer 
performances of that time. Also, in  calculations of ref. \cite{[CNR96]} 
some important orbits are missing, like some spin
orbit partners, resulting in a violation of the Ikeda sum rule of about 
50\%, which means it should be expected that they give too small M.E..

\begin{table}
\centerline{   
\begin{tabular}{|cccccccc|}
\hline
 & \cite{[SMK90]} & \cite{[TOM91]} & \cite{[HAX84]} & \cite{[CNR96]}&
 \cite{[SIM97]} & \cite{[SPV20]}& present work \\
\hline
\hline 
$M^{0\nu}$ & $4.25$ & $4.26$ & $4.85$ & $1.57$ & $1.92$ & $2.80$ &
$2.36 (12)$~~$3.62 (9)$\\
\hline
$<m_{\nu}>$ [eV] 90$\%$ & $0.345$ & $0.328$ & $0.304$ & $0.940$ & $0.768$ & $0.527$ &
$0.625$ ~~~ $0.407$\\
\hline
$<m_{\nu}>$ [eV] 68$\%$ & 0.27 &  0.26 & 0.24 & 0.73 & 0.60 & 0.41 & 0.490~~~~
0.320\\ 
\hline
\end{tabular}}
\caption{Neutrinoless M.E. and upper limits for the neutrino mass parameter 
for $^{76}Ge$, calculated with a phase space $F_1^{0\nu} = 6.31 
\times 10^{-15} ~y^{-1}$ (\cite{[SC98]})  
and $T^{0\nu}_{1/2} > 1.9 \times 10^{25}$ y (C.L. 90\%), and 
$3.1 \cdot {10}^{25} ~y$ (C.L. 68$\%$) \cite{[HM20]}. 
The non-dimensional values of various ME are taken and reconverted, 
when necessary, from the indicated references. For the present work 
two values, representing the calculation with 12  and 9 levels 
for the  s.p. basis, are displayed.}
\label{t3}
\end{table}

\section{Conclusions} 

We have performed a calculation of the two- and zero-neutrino $\beta\beta$ decay matrix 
elements 
for the case of $^{76}Ge$ with the pnQRPA, second-QRPA, RQRPA and full-RQRPA methods, using 
two different choices of the s.p. basis. We can summarize the main results as follows:

i) for the $M^{2\nu}_{GT}$ we got a significant dependence of the results on the size of 
the s.p. 
basis, in the case of the RQRPA and full-RQRPA methods, while the results obtained with 
pnQRPA and second-QRPA do not display such a dependence. 

ii) for the neutrinoless decay mode all the three methods used for the calculation, i.e. pnQRPA,
RQRPA and full-RQRPA, give differences between 9 and 12 level calculations by factors  
1.6-3. The values of the ME obtained with the three methods are close to each other for 
the calculation  with the smaller basis, while they differ significantly when the 
calculation is 
done with the larger basis. i) and ii) reveal a sensitivity of the RQRPA methods to the 
size of the s.p. basis which is used. This could have its root in the numerical 
double-iteration procedure used in RQRPA-type calculations and in our opinion further 
improvements should be done in this respect. 

iii) we also check the ISR within the four methods and found it to be 
fulfilled with a good 
approximation within second-QRPA method, while with RQRPA and full-RQRPA the deviations 
are up to 21\%. We found that this result is not much dependent on the size of the 
s.p. basis used. 
This result, besides the numerical arguments mentioned above, might also be explained by 
theoretical arguments related to the way the partial restoration of the Pauli principle 
is done within RQRPA.  The restoration is made in the commutator relations of the operators 
$A$, $A^{\dagger}$ by taking into account the 
averages of the quasiparticle-number operators in their commutator relations. However, 
there is
no justification to neglect them in the $B$, $B^{\dagger}$ operator 
commutation 
relations. However, this is done within the second-QRPA and moreover, in this method the next 
higher-order corrections beyond pnQRPA are also taken into account for the $\beta^{\pm}$ 
operators. The additional terms give a positive contribution to the ISR, while as it is 
known RQRPA underestimates the ISR.

iv) using the most recent reported neutrinoless half-life limit, and using the value of $g_{pp}$ 
fixed for the $2\nu$ neutrino mode calculation, we extracted the
following new upper limits for the neutrino mass parameter. A 
critical comparison between various values of
the M.E. found in the literature was performed. One observes a tendency
of reducing these values according to the most recent calculations
performed with RQRPA and full-RQRPA. One may conclude that the 
values of the M.E. involved in $0\nu\beta\beta$ decay of $^{76}Ge$ can be 
reliably predicted within a factor of two.
 
Finally we would like to stress that considering the various  
approximations 
made in the different calculations (violation of Ikeda sum rule in 
\cite{[CNR96]}, \cite{[SIM97]}, \cite{[SPV20]}, and this work (by 50, 20$\%$)), 
and neglection of weak magnetism and pseudoscalar coupling in all approaches 
except in \cite{[SPV20]} (another 30$\%$), the tendency goes to a variation 
within the different approaches of only a factor of 1.5, and to clearly 
favouring the smaller deduced neutrino mass values. The $< m_\nu>$ values 
expected from $^{76}{Ge}$ decay would lie around 0.2 eV (68$\%$ C.L.) after 
the corresponding estimated corrections.


\begin{thebibliography}{99}

\bibitem{[HAL67]} J. A. Halbleib and R. A. Sorensen, Nucl. Phys. {\bf A 98} 
(1967) 542.

\bibitem{[HAX84]} W. C. Haxton and G. J. Stephenson, Progr. Part. Nucl. Phys.
{\bf 12} (1984) 409.

\bibitem{[GK86]} H.V. Klapdor and K. Grotz, Phys. Lett. {\bf B 142} (1984) 
323; K. Grotz and H.V. Klapdor, Phys. Lett. {\bf B 153} (1985) 1;
{\bf B 157} (1985) 242; Nucl. Phys. {\bf A 460} (1986) 395.


\bibitem{[VOG86]} P. Vogel and M. R. Zirnbauer, Phys. Rev. Lett.
{\bf 57} (1986) 3148.

\bibitem{[CFT87]} O. Civitarese, A. Faessler and T. Tomoda, Phys. Lett. {\bf B 194} (1987) 11; T. Tomoda and A. Faessler, Phys. Lett. {\bf B 199} (1987) 475.

\bibitem{[SUH88]} J. Suhonen, A. Faessler, T. Taigel and T. Tomoda, Phys. Lett.
{\bf B 202} (1988) 174.

\bibitem{[STAU90]} A. Staudt, T.T.S. Kuo and H.V. Klapdor-Kleingrothaus, Phys. 
Lett. {\bf B 242} (1990) 17.

\bibitem{[SMK90]} A. Staudt, K. Muto and H.V. Klapdor-Kleingrothaus, Europhys. Lett. {\bf 13} (1990) 31.

\bibitem{[RAD91]} A. A. Raduta, A. Faessler, S. Stoica and W. A. Kaminski,\\
Phys. Lett. {\bf B 254} (1991) 7; A.A. Raduta, A. Faessler and S. Stoica, Nucl. Phys. 
{\bf A 534} (1991) 149.

\bibitem{[CFS91]} O. Civitarese, A. Faessler, J. Suhonen and X.R. Wu, J. Phys. 
{\bf G 17} (1991) 943.

\bibitem{[TOM91]} T. Tomoda, Rep. Prog. Phys. {\bf 54} (1991) 53.

\bibitem{[GV92]} A. Griffits and P. Vogel, Phys. Rev. {\bf C 46} (1992) 181.


\bibitem{[KRM93]} F. Krmpotic, A. Mariano, T.T.S. Kuo, and N. Nakayama, Phys. Lett. {\bf B 319} 
(1993) 393. 

\bibitem{[SUH93]} J. Suhonen, Nucl. Phys. {\bf A 563} (1993) 205; J. Suhonen and O. Civitarese, 
Phys. Lett. {\bf B 308} (1993) 212. 

\bibitem{[STO94]} S. Stoica and W.A. Kaminski, Phys. Rev. {\bf C 47} (1993) 867; 
S. Stoica, \\ Phys. Rev. {\bf C 49} (1994) 787.

\bibitem{[STO95]} S. Stoica, Phys. Lett. {\bf B 350} (1995) 152. 

\bibitem{[RDF]} A.A. Raduta, D.S. Delion and A. Faessler, Phys. Lett. {\bf B 312} (1993) 13;
Phys. Rev. {\bf C 51} (1995) 3008.

\bibitem{[TOI95]} J. Toivanen and J. Suhonen, Phys. Rev. Lett. 
{\bf 75} (1995) 410; Phys. Rev. {\bf C 55} (1997) 2314. 

\bibitem{[CHE95]} M.K. Cheoun, A. Faessler, F. Simcovic, G. Teneva and A. Bobyk, Nucl. Phys. 
{\bf A 587} (1995) 301.

\bibitem{[SSF96]} J. Schwieger, F. Simcovic and A. Faessler, Nucl. Phys. {\bf A 600} (1996) 179.

\bibitem{[PAN96]} G. Pantis, F. Simcovic, J.D. Vergados and A. Faessler, 
Phys. Rev. {\bf C 53} (1996) 695.


\bibitem{[RS96]} A.A. Raduta and J. Suhonen, Phys. Rev. {\bf C 53} (1996) 176; J. Phys. 
{\bf G 22} (1996) 123.



\bibitem{[CHH97]} O. Civitarese, P.O. Hess and J.G. Hirsch, P.O. Hess,
Phys. Lett. {\bf B 412} (1997) 1; J.G. Hirsch, P.O.Hess and O. Civitarese,
Phys. Rev. {\bf C 54} (1996) 1976. 

\bibitem{[SCW97]} J. Schwieger, F. Simcovic, A. Faessler, and W.A. Kaminski, J. Phys. {\bf G 23} (1997) 1647; Phys. Rev. {\bf C 57} (1998) 1738.

\bibitem{[SIM97]} F. Simkovic, J. Schwieger, M. Veselsky, G. Pantis and A. Faessler, Phys. Lett. {\bf 393} (1997) 267.

\bibitem{[SC98]} J. Suhonen and O. Civitarese, Phys. Rep. {\bf 300} (1998) 123.

\bibitem{[BKZ99]} A. Bobyk, W.A. Kaminski and P. Zareba, Eur. Phys. J. {\bf A 5} (1999) 
385; \\ Nucl. Phys. {\bf  A 669} (2000) 221.

\bibitem{[CNR96]} E. Caurier, F. Nowacki, A. Poves and J. Retamosa, Phys. Rev. Lett. {\bf 77} 
(1996) 1954. 

\bibitem{[HM99]} Heidelberg-Moscow collaboration: L. Baudis et al., Phys. 
Lett. {\bf B 407} (1997) 219; Phys. Rev. Lett. {\bf 83} (1999) 41; 

\bibitem{[HM20]}Heidelberg-Moscow collaboration (submitted to Phys. Lett. 
{\bf B}) (2000);


\bibitem{[BZ62]} S.T. Beliaev and G. Zelevinski, Nucl. Phys. {\bf 39} (1962) 582.

\bibitem{[HAR64]} K. Hara, Prog. Theor. Phys. {\bf 32} (1964) 88.

\bibitem{[IKE65]} K. Ikeda, T. Udagawa and H. Yamamura, Prog. Theor. Phys. {\bf 33} (1965) 22. 

\bibitem{[ROW68]} D. J. Rowe, Rev. Mod. Phys. {\bf 40} (1968) 153; 
Nucl. Phys. {\bf A107} (1968) 99.


\bibitem{[SPV20]} F. Simkovic, G. Pantis, J.D. Vergados and A. Faessler,
hep-ph/9905509. 

\bibitem{[FAE99]} A. Faessler and F. Simkovic, hep-ph/9901215.

\end{thebibliography}
\end{document}